\documentclass[12pt]{article}
\newtheorem{lemma}{Lemma}[section]
\newtheorem{theorem}[lemma]{Theorem}
\newtheorem{proposition}[lemma]{Proposition}

\newtheorem{remark}[lemma]{Remark}

\newtheorem{definition}[lemma]{Definition}

\newcommand{\R}{{\rm I\mkern-4.0mu R}}
\def\C{ {\rm C \kern -.15cm \vrule width.5pt \kern .12cm}}
\def\Z{ {\rm Z \kern -.27cm \angle \kern .02cm}}
\def\N{ {\rm N \kern -.26cm \vrule width.4pt \kern .10cm}}
\def\1{{\rm 1\mskip-4.5mu l} }
\def\lsim{\raise0.3ex\hbox{$<$\kern-0.75em\raise-1.1ex\hbox{$\sim$}}}
\def\gsim{\raise0.3ex\hbox{$>$\kern-0.75em\raise-1.1ex\hbox{$\sim$}}}
\def\noi{\noindent}

\def\beq{\begin{equation}}   \def\eeq{\end{equation}}
\def\bea{\begin{eqnarray}}  \def\eea{\end{eqnarray}}
\def\nn{\nonumber}
\def\noi{\noindent}
\newcommand{\QED}{\mbox{}\hfill
\raisebox{-2pt}{\rule{5.6pt}{8pt}\rule{4pt}{0pt}}
            \medskip\par}
\newcommand\mysection{\setcounter{equation}{0}\section}
\renewcommand{\theequation}{\thesection.\arabic{equation}}
\newcounter{hran} \renewcommand{\thehran}{\thesection.\arabic{hran}}

\def\bmini{\setcounter{hran}{\value{equation}}
    \refstepcounter{hran}\setcounter{equation}{0}
    \renewcommand{\theequation}{\thehran\alph{equation}}\begin{eqnarray}}

\def\bminiG#1{\setcounter{hran}{\value{equation}}
\refstepcounter{hran}\setcounter{equation}{-1}
\renewcommand{\theequation}{\thehran\alph{equation}}
\refstepcounter{equation}\label{#1}\begin{eqnarray}}

\def\emini{\end{eqnarray}\relax\setcounter{equation}{\value{hran}}\renewcommand{
\theequation}
{\thesection.\arabic{equation}}}

\begin{document}
\title {Semiclassical results in the linear response theory}
\author{\it {\bf Monique Combescure} \\
\it Laboratoire de Physique Th\'eorique,   CNRS - UMR 8627\\
\it Universit\'e de Paris XI, B\^atiment
210, F-91405 ORSAY Cedex, France \\
\it and \\
\it IPNL, B\^atiment Paul Dirac \\
\it 4 rue Enrico Fermi,Universit\'e Lyon-1 \\
\it  F.69622 VILLEURBANNE Cedex, France\\
\it email monique.combescure@ipnl.in2p3.fr\\
\\
\it {\bf Didier Robert} \\
\it D\'epartement de Math\'ematiques\\
\it Laboratoire Jean Leray, CNRS-UMR 6629\\
\it Universit\'e de Nantes, 2 rue de la Houssini\`ere, \\
F-44322 NANTES Cedex 03, France\\
\it email Didier.Robert@math.univ-nantes.fr}
\vskip 1 truecm
\date{}
\maketitle

\begin{abstract}
We consider a quantum system of non-interacting fermions at
temperature T, in the framework
   of linear response theory. We show that semiclassical theory is an
appropriate framework to
   describe some of their thermodynamic properties, in particular
through asymptotic expansions
    in $\hbar$ (Planck constant) of the dynamical susceptibilities. We
show how the closed orbits of
    the classical motion in phase space manifest themselves in these
expansions, in the regime
     where T is of the order of $\hbar$.
\end{abstract}

\vskip 1.5 truecm

\newpage
\pagestyle{myheadings}
\baselineskip 18pt
\mysection{Introduction}
\hspace*{\parindent} Consider a system of non-interacting fermions
confined by an
external potential and in contact with an exterior reservoir at
temperature $T$.
Assume that a time-varying external perturbation drives the system
out to, but near of,
its equilibrium state. The response of this quantum system to an external
time-dependent perturbation is a subject of high physical interest,
which can be
investigated experimentally, in particular the so-called ``dynamical
susceptibility''.
A complete rigorous analysis of this problem is still lacking, although recent
progress is being made in the understanding of non-equilibrium
statistical mechanism,
and its link with the underlying chaotic dynamics \cite{GC,GR,R}. \par

A semi-empirical route which has been proposed (see classical textbooks
\cite{KTH,LL}) consists, for small perturbation, of investigating the response
function ``to first order of the perturbation'', i.e. the so-called ``linear
response theory''. This semi-empirical route has been given a firmer
foundation (see
the book by Bratelli and Robinson \cite{BR2}) where a link with the
KMS condition is
established. (See also recent progress in \cite{R}). \par

In this paper we rederive  the first order response function for the
quantum fermionic system under study, i.e. the so-called ``generalized Kubo
formula'' (see also \cite{AG}) and investigate semiclassical expansions of it,
assuming suitable ``chaoticity assumptions'' on the one-body
underlying classical
dynamics. These semiclassical expansions are developed in a similar spirit as
previous studies on the ``semiclassical magnetic response for non-interacting
electrons'' \cite{A,B,CR,F,GJ,MR,RUJ} i.e. we exhibit a low temperature
regime where the closed classical orbits of one-particle motion manifest
themselves as oscillating corrections to the response function. \par

Section 2 contains , for pedagogical purposes the basic framework of so-called
``second quantization'' in which the physical system under consideration can be
studied and its thermodynamical properties mathematically investigated.
Section 3 presents the so-called ``linear-response theory'', and the
dynamical susceptibility
that will be studied in the semiclassical framework. Section 4
presents and derives the
main results of this paper: a rigorous semiclassical expansion of the
dynamical susceptibility
under suitable assumptions on the physical system.

\mysection{The physical model}
\hspace*{\parindent} Consider a system of non-interacting fermions,
living in ${I\hskip-1truemm
R}^n$, subject to a one-body Hamiltonian $\widehat{H}$ which is the Weyl
quantization of a classical Hamiltonian $H(q, p)$ of the form

\beq
\label{2.1e}
H(q,p) = {p^2 \over 2m} + V(q)
\eeq

\noi with $V \in {\cal C}^{\infty} (\R^n)$ such that
the following confining  assumption holds:\\
\\
\textbf{Assumption 1}:
\\
$$V(q) \geq c_0 (1 + q^2)^{s/2} \qquad s\ , c_0 > 0$$
\\

Under these assumptions, $\widehat{H}$ is self-adjoint in $L^2({I\hskip-1truemm
R}^n)=\mathcal{H}$ and its spectrum is pure point, and contained in
$]0, \infty )$.\par
\medskip

    Assume that the system of non-interacting fermions
is infinite and in contact with a reservoir at temperature $T$. The
study of thermodynamical
properties of this system is performed within the framework of
statistical mechanics which is
   well known, and that we recall here for completeness (see \cite{B}).
We introduce the so-called
   Fock space:

   \beq
   \label{2.2e}
   {\mathcal{F}}_a ={\bigoplus }_{n=0}^{\infty}\left(\otimes_a^n
\mathcal{H}\right)
   \eeq
  where
  $\otimes_a^n \mathcal{H}$ is the antisymmetric tensor product of
   $\mathcal{H}$,  which physically
   represents the space of n-fermions states. The Hamiltonian of the
infinite system is governed by
   the second quantization of $\widehat{H}$:

   \beq
   \label{2.3e}
   d\Gamma(\widehat{H})=\widehat{\widehat{H}}
   \eeq

   \noi acting in ${\mathcal{F}}_a$. Similarly the number
$\widehat{\widehat{N}}$ of particles
   is a second quantized operator in  ${\mathcal{F}}_{a}$:

   \beq
   \label{2.4e}
   \widehat{\widehat{H}}=d\Gamma(\1_{\mathcal{H}})
   \eeq

   Note that if $\left\{\psi_j\right\}_{j\geq 0}$ denotes an
orthonormal basis of
   eigenfunctions of $\widehat{H}$, with eigenvalue $E_j$:

   \beq
   \label{2.5e}
   \widehat{H}{\psi}_j=E_j{\psi}_j
   \eeq

   \noi then:

   \beq
   \label{2.6e}
   \left\{{\psi}_{j_1} \wedge{\psi}_{j_2} \wedge...\wedge{\psi}_{j_n}
\right\}_{j_1<j_2<...<j_n}
   \eeq

    \noi is an orthonormal basis of ${\otimes}_a^n{\mathcal{H}}$
consisting of eigenvectors
    of $\widehat{\widehat{H}}$ with eigenvalue
    $E_{j_1}+...+E_{j_n}$.

\noi  According to the Pauli principle, the occupation number $n_j$ of
   any  state ${\psi}_j$
   in ${\mathcal{F}}_a$ equals 0 or 1. Thus the spectrum of
$\widehat{\widehat{H}}$ can
   be rewritten as:

   $${\sum}_{j}n_{j}E_{j}$$

   \noi where $n_j$ is the eigenvalue of $\widehat{N}_{j,n}$:

   \beq
   \label{2.7e}
   \widehat{N}_{j,n}({\psi}_{j_1} \wedge{\psi}_{j_2}
\wedge...\wedge{\psi}_{j_n})=n_{j}({\psi}_{j_1}
    \wedge{\psi}_{j_2} \wedge...\wedge{\psi}_{j_n})
   \eeq

   \noi ($\widehat{N}_{j,n}$ ``tells'' whether or not the state
${\psi}_j$ is occupied in a given
   state of ${\otimes}_a^n\mathcal{H}$). We define:

   $${\widehat{N}}_j={\oplus}_{n\geq0}\widehat{N}_{j,n}$$

   Obviously we have:

   \bea
   \label{2.8e}
   \widehat{\widehat{N}}={\sum}_{j\geq1}\widehat{N_j}\\
   \widehat{\widehat{H}}={\sum}_{j\geq1}\widehat{N_j}E_j
   \nonumber
   \eea

   \noi Note that: $[\widehat{\widehat{H}},\widehat{\widehat{N}}]=0$

\noi In the grand-canonical formalism (see \cite{B}), the Gibbs
partition function is:

   \beq
   \label{2.9e}
   Z_G= {\rm Tr}\left(
e^{-\beta\widehat{\widehat{H}}+\kappa\widehat{\widehat{N}}}\right)
   \eeq

   \noi where $\kappa$ and $\beta$ are Lagrange multipliers:

   \bea
   \label{2.10e}
   \beta= 1/kT\\
   \kappa=\beta\mu
   \nn
   \eea

   \noi $\mu$ being the chemical potential, and the Trace (which we
denote with capital T)
   being taken in ${\mathcal{F}}_a$. Then it
   can easily be shown (see \cite{B}) that $Z_G$ factorizes as:

   \beq
   \label{2.11e}
   Z_G ={\prod}_{j\geq1} \left(1+e^{-\beta(E_{j}-\mu)}\right)
   \eeq

\noi The mean value $F_j$ of the occupation number of ${\psi}_j$ is then:

   \bea
   \label{2.12e}
   F_j = {Z_G}^{-1} {\rm Tr} \left(
\widehat{N_j}e^{-\beta(\widehat{\widehat{H}}-\widehat{\widehat{N}})}
   \right)\\
   ={\left( 1+e^{\beta(E_{j}-\mu)}\right)}^{-1}
   \nn
   \eea

   \noi Denoting by $f$ the Fermi-Dirac function:

   \beq
   \label{2.13e}
   f(x) = (1+ e^x)^{-1}
   \eeq

   \noi the mean value of the number of particles in the
grand-canonical ensemble is then:

   \beq
   <N> = {\sum}_{j\geq1} F_j = {\rm tr}
\left\{f\left(\beta(\widehat{H}-\mu)\right)\right\}
   \eeq

   \noi where now the trace (which we here denote with small t) is
taken in $\mathcal{H}$.
    The operator in $\mathcal{H}$:

   \beq
   \label{2.14e}
   \widehat{\rho}_{eq} := f\left(\beta(\widehat{H}-\mu)\right)
   \eeq

   \noi is called the Fermi-Dirac equilibrium one-body operator.\\
   \\
   We now assume that the one-body Hamiltonian is slightly perturbed in
a time-dependent way:

   \beq
   \label{2.15e}
   {\widehat{H}}_{\lambda}(t) = \widehat{H} + \lambda \widehat{A} F(t)
   \eeq

   \noi where $\lambda$ is a small real parameter, and $F$ of the form,
with $\alpha>0$ :
\beq
\label{2.16e}
F(t) = \cases{e^{\alpha t} \qquad t < 0 \cr 1 \qquad \ \  t \geq 0 \ . \cr}
\eeq

   \noi Starting at time $t = -\infty$ from the Fermi-Dirac one-body
equilibrium state
   $\widehat{\rho}_{eq}$, and switching on the pertubation , we get
   a time-dependent ``density matrix'' $\widehat{\rho}_{\lambda}(t)$
(namely a trace one operator)
   obeying:

   \bea
   \label{2.17e}
   i \hbar {\partial \widehat{\rho}_{\lambda} \over \partial t}
    = \left [ \widehat{H_{\lambda}}(t), \widehat{\rho}_{\lambda} \right ]\\
    \lim_{t \to - \infty}\widehat{ \rho_{ \lambda}} (t) =\widehat{ \rho_{eq}}
    \nn
    \eea

    \noi The PROBLEM is the following: to which extend does
$\widehat{\rho}_{\lambda}(t)$
    wander from the equilibrium state $\widehat{\rho}_{eq}$ as the
perturbation is switched on?

   \mysection{The linear response theory}

   Physically, we aim to answer the above PROBLEM ``to the first order
in $\lambda$'', whence
   the name ``linear response theory''. In this section we give a
rigorous framework to this program.
   Thus our first step is to set a convenient set of assumptions on the
Hamiltonians under which
   mathematical results can be obtained.
   \bigskip

  \noi \textbf{Assumption 2}
   \\
   $A(q)$ is a multiplicative function dominated by $C(1+q^2)$ in
absolute value.\\

   Under this assumption we know (\cite{Y}) that the unitary evolution
operator generated by
     $\widehat{H}_{\lambda}(t)$, namely solving:

   \bea
   \label{3.1e}
   i\hbar {\partial V_{\lambda}(t, t') \over \partial t} =
   \widehat{H_{\lambda}}(t)\ V_{\lambda}(t, t')\\
   V_{\lambda}(t_0, t_0) = \1
   \nn
   \eea
  exists. Moreover it obeys the Duhamel's formula:
\beq
   \label{3.2e}
   V_{\lambda}(t, t_0) = U(t - t_0) + {\lambda \over i \hbar}
\int_{t_0}^t dt' \ V_{\lambda}(t , t') \ F(t') \  \widehat{A}
U(t^\prime - t_0)
   \eeq
where we have denoted:

   \beq
   \label{3.3e}
   U(t) : = e^{-it\widehat{H}/\hbar}.
   \eeq

   We define:

   \beq
   \label{3.4e}
   \widehat{\rho}_{\lambda} (t,t_0) = V(t, t_0) \widehat{ \rho}_{eq}\ V(t_0, t)
\eeq

   \noi It is clearly a solution of (\ref{2.17e}) with $\widehat{\rho
}_{\lambda}(t_0 ) =
   \widehat{\rho}_{eq}$.

We shall now justify the {\bf ``linear response theory''}  in this context.

\begin{proposition}

The mapping $\lambda \mapsto \widehat{\rho}_{\lambda}(t, t_{0})$ given by
(\ref{3.4e})
is differentiable near $\lambda = 0$ in the trace-class operator norm
sense and we
have~:

\beq
\label{3.5e}
\left . {d \over d \lambda} \widehat{\rho}_{\lambda}(t, t_0) \right
|_{\lambda = 0} = {1 \over i
\hbar} \int_{t_0}^t dt'\ F(t') \ U(t - t') [\widehat{\rho}_{eq} ,
\widehat{A}] U(t' - t) \ .
\eeq

\noi Moreover  $\widehat{\rho}_{\lambda} (t,t_0)$ has a limit as $t_0
\to - \infty$, in the
trace-class operator norm sense~,  called
$\widehat{\rho}_{eq}(t,\lambda)$ and which is also differentiable in
$\lambda$.  Moreover  we have:
\beq
\left . {d \over d \lambda} \widehat{\rho}_{\lambda}(t) \right
|_{\lambda = 0} = {1 \over i
\hbar} \int_{-\infty}^t ds\ F(s) [\widehat{\rho}_{eq} ,
{\widehat{A}}_{t-s}],
\eeq
\noi $\widehat{A}_t$ being, by definition, the Heisenberg observable at
time $t$  (for the quantum evolution governed by $\widehat{H}$):
\beq
\widehat{A}_t = U(t) \ \widehat{A} \ U(t)^* \ .
\eeq

\end{proposition}

   {\bf Proof}: Inserting  $\1=U(t_0-t) U(t-t_0)$ and commuting with $
\widehat{\rho}_{eq}^{1/2}$ we
obtain ~:

\beq
\label{3.6e}
   \widehat{\rho}_{\lambda}
(t,t_0)=V_{\lambda}(t,t_0)U(t_0-t)\widehat\rho_{eq}^{1/2} .
\widehat\rho_{eq}^{1/2} U(t-t_0)V_{\lambda}(t_0,t)
\eeq

\noi Each of these two factors admits a limit as $t_0\to- \infty$ in
the norm trace sense. Namely using Duhamel's formula, we have~:

\beq
\label{3.7e}
\left(V_{\lambda}(t,t_0)U(t_0-t)-\1\right)\widehat{\rho}_{eq}^{1/2}=\frac{\lambda}
{i\hbar}
\int_{t_0}^t F(s)V_{\lambda}(t,s)\widehat{A}\widehat{\rho}_{eq}^{1/2} U(s-t)ds
\eeq

\noi (and similarly for the adjoint)  so the result follows since
$\int_{-\infty}^t \vert F(s) \vert ds $ exists for any finite $t$. \par

Letting $t_0$ tend to $-\infty$, we then get:

\beq
\label{3.8e}
\widehat{\rho}_{eq}(t,\lambda):=\widehat{\rho_{eq}}-\frac{\lambda}{i\hbar}
\int_{-\infty}^t dsF(s)
V_{\lambda}(t,s) [\widehat{A}, \widehat{\rho}_{eq}] V_{\lambda}(s,t)
\eeq

A Taylor expansion near $\lambda =0$ of $\widehat{\rho}_{eq}(t, \lambda)$
can be obtained by plugging in
   the Duhamel's formula  in (\ref{3.8e}) :

   \beq
V_{\lambda}(t,s)= U(t-s)+\frac{\lambda}{i\hbar} \int_{s}^t
F(\sigma)V_{\lambda}(t,\sigma) \widehat{A}
U(\sigma-t) d\sigma
\eeq

This gives in the trace norm sense:

\beq
\widehat{\rho}_{eq}(t,\lambda)-\widehat{\rho}_{eq}= \frac{\lambda}{i\hbar}
\int_{-\infty}^t dt' F(t') [\widehat{\rho}_{eq} , \widehat{A}_{t-t'}]
+o (\lambda).
\eeq
By this method the second term in $\lambda$  (``quadratic response'') could
also  be explicitly written.
\QED

\noi Equation (\ref{3.8e}) is the linear response formula in this framework. It
implies that if $\widehat{B}$ is some self-adjoint operator that we
want to measure
in the ``almost   stationary'' state $\widehat{\rho}_{eq} (t)$, the
coefficient of the first
order contribution in $\lambda$,
as $\lambda \to 0$  to the result~:
\beq
\label{3.9e}
J_{\lambda}(t) = {\rm tr}\left \{ \widehat{B} \left (\widehat{ \rho}_{eq}
(t, \lambda) -
\widehat{ \rho}_{eq}\right ) \right \}
\eeq

\noi is of the form :

\beq
\label{3.10e}
J_L(t) = \lambda\int_{-\infty}^t dt'\ F(t') \ \Phi (t - t')
\eeq

\noi where

\beq
\label{3.11e}
\Phi (t)  =  {1\over i \hbar} {\rm tr} \left ( \widehat{B} \left [\widehat{
\rho}_{eq} , \widehat{A}_t
\right ] \right )
   =  {1 \over i \hbar} {\rm tr} \left ( \widehat{\rho}_{eq} \left [
\widehat{A} ,\widehat{B}_{-t} \right ] \right )
\nn
   \eeq
\noi using the cyclicity of the trace. \\  \noi

We now take the Fourier transform, in the distributional sense of
$\Phi (t)$, called
the ``generalized susceptiblity''~:

\beq
\label{3.12e}
\chi_{A,B}(\omega ) = \int_{-\infty}^{+ \infty} \Phi (t)\ e^{i\omega t}\ dt
\eeq

\noi which is the quantity that we shall
   study now.
Given any function $g$ whose Fourier Transform $\widetilde{g}$ is
assumed to belong to
${\cal C}_0^{\infty}(\R)$~:

\beq
\label{3.13e}
\int\chi_{A,B}(\omega)g(\omega)
d\omega={1\over{i\hbar}}\int_{-\infty}^{+\infty}
{\rm tr} f(\beta(\widehat{H}-\mu))
[\widehat{A} ,\widehat{B}_{-t}] \widetilde{g}(-t) dt :=I(\mu)
\eeq

\noi Our aim is to obtain a semiclassical expansion of
$\chi_{A,B}(\omega)$ as $\hbar\to 0$,
   $\beta \to\infty$, namely a semiclassical expansion at low temperature.

It is useful to introduce the following parameter:

\beq
\label{3.14e}
\sigma = \beta \hbar
\eeq

\noi which has the dimension of time. We also define the function
$f_{\sigma}$ as follows:

\beq
\label{3.15e}
f_{\sigma}(x) = (1+e^{\sigma x})^{-1}
\eeq

\noi so that $I(\mu)$ can be rewritten formally as:

\beq
\label{3.16e}
I(\mu) = {1\over{i\hbar}}\int {\rm tr}
\left(f_{\sigma}\left(\frac{\widehat{H}-\mu}{\hbar}\right)
[ \widehat{A} ,\widehat{B}_t] \right) \widetilde{g}(t) dt
\eeq

However this expression suffers from the singularity in 0 of
$f_{\sigma}$ as $\sigma\to0$.
In order to avoid this, we ``regularize'' it by using instead of $f_{\sigma}$:

\beq
\label{3.17e}
f_{\sigma, \eta} = f_{\sigma}* \eta
\eeq
\noi where $\eta$ is a function in $\mathcal{S}(\R)$ such that its
Fourier Transform
$\widetilde{\eta} \in {\mathcal{C}}_0^{\infty}(\R)$.
This amounts to study ${\chi}_{A, B}$ as a distribution on ${\R}^2$
in the variables $s$ and
$\omega$ in the following way:
\bea
\label{3.18e}
\int\int\chi_{A,B}(s,\omega)\widetilde{\eta}(s)g(\omega)dsd\omega=\\
\frac{1}{2i\pi\hbar}\int\int {\rm tr}\left(e^{is(\widehat{H}-\mu)/{\hbar}}
\left[\widehat{A},
{\widehat{B}}_t\right]\right)\widetilde{f_\sigma}(s)\widetilde\eta(s)
\widetilde{g}(t)dsdt
\nn
\eea
\\
Let us intoduce the following test space functions on $ \R^2 =
\R_s\times\R_\omega$:

\begin{definition}\label{test}
We say that  $\varphi \in {\mathcal K}_a$, $a>0$,  if $\varphi$ is
$C^\infty$ on
    $\R^2$ and there exist $b>0$, $c>0$ such that $\varphi(s, \omega) = 0$
    for $\vert s\vert \geq b$, $\omega\in\R$, and
    $\vert\tilde{\varphi}^{(2)}(s, t)\vert \leq c{\rm e}^{-a\vert t\vert}$
    for every $(s, t)\in\R^2$, where $\tilde{\varphi}^{(2)}(s, t)$ denotes
the Fourier
    transform in the second argument.
\end{definition}

\mysection{The results}

In this section we first introduce the notations of the classical
objects that will appear
in the  semiclassical expansions, together with the assumptions under which
these expansions can be obtained.
\medskip

Let $\phi^t$ be the classical flow induced by Hamiltonian
(\ref{2.1e}). Consider $\Sigma_\mu$
   the energy surface conserved by the flow:

\beq
\label{4.1e}
\Sigma_\mu=\left\{ (q,p)\in \R^{2n} :H(q,p)=\mu \right\}
\eeq

\noi We call $d\Sigma_{\mu}$ the Liouville measure on $\Sigma_\mu$,
so that the correlation of
classical observables $A$ and $B$ on  $\Sigma_\mu$ is defined by:

\beq
\label{4.2e}
C_{A,B, \mu}(t)=\int_{\Sigma_\mu}A.B_{t}d\Sigma_\mu
\eeq

\noi where $B_{t}(z)=B[\phi^{t}(z)]$. Moreover if $\gamma$ is any
periodic orbit on $\Sigma_\mu$
, and $\gamma*$ the corresponding primitive orbit, with period
$T_{\gamma*}$, we introduce
the correlation function

\beq
\label{4.3e}
c_{\gamma*}(t)=\int_0^{T_{\gamma*}} A_{s}(q,p)B_{s+t}(q,p)ds  \qquad
(q,p)\in\gamma*
\eeq

$c_{\gamma*}$ being $T_{\gamma*}$-periodic it admits the
Fourier-series expansion:

\beq
\label{4.4e}
c_{\gamma*}(t)=\sum_{k=-\infty}^{k=+\infty}
c_{\gamma*,k}e^{2ikt\pi/T_{\gamma*}}
\eeq

To each $\gamma$ is associated a corresponding ``linearized Poincar\'e
map'' called $P_{\gamma}$,
a classical action along $\gamma$ called $S_{\gamma}$, and a Maslov
index $\nu_{\gamma}$
(see \cite{CRR} ).
\\
Let us assume that $\phi^t$ on $\Sigma_{\mu}$ satisfies the so-called
Gutzwiller Assumption:\\
\medskip
\noi
\textbf{Assumption 2} The periodic orbits $\gamma$ are
non-degenerate, i.e the Poincar\'e maps
do not have 1 as eigenvalue (which implies that they are isolated).\\
\medskip
\noi Moreover we shall be able to treat $B$ obeying:\\
\bigskip
\textbf{Assumption 3}
$$\vert  \partial_q^{\alpha}\partial_p^{\beta} B(q,p)\vert\leq
C_{\alpha \beta}\qquad
\vert\alpha\vert + \vert\beta\vert\geq 2$$

\noi Our result is as follows:

\begin{theorem}
Under {\rm \textbf{Assumptions 1,2,3}}, we have, in distributional
   sense in ${\mathcal K}_a$ (see definition \ref{test}) :
$$\chi_{A,B}(s,\omega) =
-{h}^{-n}\delta_0(s)\otimes\widetilde{C'_{A,B, \mu}}(\omega)
+\sum_{j\geq1} {\hbar}^{j-n} \mu_j(s,\omega)\qquad$$

$$+\sum_{\gamma :T_{\gamma}\not=0}\frac{\pi
e^{i(S_{\gamma}/\hbar+\nu_{\gamma}\pi/2)}}
{\hbar\sigma \sinh {(\pi T_{\gamma}/{\sigma})}\vert
det(1-P_{\gamma})\vert ^{1/2}}
\left(\delta_{T_{\gamma}}(s)
\otimes\sum_{k}c_{\gamma*,k}\delta(\omega-\frac{2k\pi}{T_{\gamma*}}
) +\sum_{j\geq1} {\hbar}^j\nu_{j,\gamma}(s,\omega)\right)$$
$$+ O(\hbar^{a\gamma_H -\varepsilon -n})$$
where $\mu_j$ and $\nu_{j,\gamma}$ are distributions in  ${\mathcal K}_a$
such that $Supp(\mu_j)\subseteq\{0\}\times \R$,
$Supp(\nu_{j,\gamma})\subseteq \{T_{\gamma}\}\times\R$, and $\gamma_H$ is a non
negative constant
    depending only on $H$ and $\mu$ (not on $a$).

\end{theorem}

{\bf Proof}:

As a distribution acting on $(\widetilde{\eta}\otimes g)(s, \omega)$,
${\chi}_{A, B}$ is given by
(\ref {3.18e}). We split the integral over t into two parts:
$\vert t \vert < {\gamma}_H Log(1/\hbar)$
   and its complement, where ${\gamma}_H$ is a constant obtained in
Egorov-type estimates (see \cite
   {BR1}) and only depending  on Hamiltonian $H$.\\
   \noi Using the exponential decrease of $\widetilde{g}(t)$, it is not
difficult to estimate the
   contribution of the integration domain $\vert t \vert > {\gamma}_H
Log(1/\hbar)$ as
   $ O(\hbar^{a\gamma_H -\varepsilon -n})$, for any $\varepsilon >0$.
The larger is $a$ (the
    exponential fall-off rate of $\widetilde{g}$) the smaller is this
``error term''.

\noi In order to estimate the contribution of the integration domain
   $\vert t \vert < {\gamma}_H Log(1/\hbar)$ we shall use truncations
in the spectral variable of
   Hamiltonian $\widehat{H}$ in
   order to apply known results and usual  methods.\\
   In all that follows, the integration support in $t$ variable is supposed
to be
   $\vert t \vert < {\gamma}_H Log(1/\hbar)$, and we call $I_{\eta} (\mu)$ the
   resulting contribution to (\ref{3.18e}).
    Fix $\delta$ positive and small enough and let
   us introduce a   ${\cal C}^{\infty}$ partition of unity as follows~:

\beq
\label {4.6e}
1=\zeta_{-}+\zeta_{0}+\zeta_{+}
\eeq

\noi where

\beq
\label{4.7e}
\zeta_{0}(t) = \cases{1 \qquad \vert{t}\vert\leq\delta/2\cr 0 \qquad
\ \ \vert{ t}\vert \geq\delta  . \cr}
\eeq
  and  Supp$\zeta_{-}\subseteq]-\infty, -\delta/2]$,
   Supp$\zeta_{+}\subseteq[\delta/2, +\infty[$.

\noi Inserting in (\ref{3.18e})\\

$$\1=\zeta_{-}(\widehat{H}-\mu)+\zeta_{0}(\widehat{H}-\mu)+\zeta_{+}(\widehat{H}
-\mu)$$
we obtain, with obvious notations~:

\beq
\label{4.8e}
   I_{\eta}(\mu) = I_{\eta}^{0}(\mu) + I_{\eta}^{+}(\mu) + I_{\eta}^{-}(\mu)
\eeq

\noi Let $\theta$ be a regular Schwartz function such that its
Fourier Transform
$\widetilde{\theta}$  be in ${\cal C}_0^{\infty}(\R)$, and

   \beq
\label{4.9e}
\widetilde{\theta} (t) \equiv \cases{1 \qquad \hbox{if}\ |t| \leq 1 \cr
0 \qquad \hbox{if}\ |t| \geq 2 \cr
}
\eeq

\noi For any positive number $\tau$, we set~:

\beq
\label{4.10e}
{\widetilde{\theta}}_{\tau} (s) := \widetilde{{\theta}} (s/\tau ) \ .
\eeq

\noi and let us denote  by ${\theta}_{\tau}$ the inverse
Fourier transform
of $ {\widetilde{\theta}}_{\tau}$. Let ${\tau}_0$ be a positive
number, small enough
in a sense to be made precise later. We shall now decompose
$I_{\eta}^{0}(\mu)$ in two parts:

\bea
\label{4.11e}
I_{\eta}^0(\mu)=I_{\eta, {\tau}_0}^0(\mu) + \nonumber\\
{1\over2i\pi\hbar}{\rm tr }
\left[\int\int ds dt
\widetilde{f}_{\sigma}(s)\widetilde{\eta}(s)\left(1-\widetilde{{\theta}}(s/\tau_
0)\right)
e^{is(\widehat{H}-\mu)/\hbar}\left[\widehat{A},\widehat{B}_t\right]
{\zeta}_0(\widehat{H}-\mu)
\widetilde{g}(t)\right]
\eea
\noi
Thus  equ. (\ref{4.8e}) now becomes:

\beq
\label{4.12e}
I_{\eta}(\mu) = I_{\eta, {\tau}_0}^{0}(\mu) +I_{\eta,
{\tau}_0}^{osc}(\mu)+ I_{\eta}^{+}(\mu)
   + I_{\eta}^{-}(\mu)
\eeq

\noi where each term can be estimated separately.
\bigskip

\noi \textbf{Estimate of $I_{\eta}^{-}(\mu)$}~:

Denote by~:
\\
$\phi_{\beta}(E):=f(\beta(E-\mu))\zeta_{-}(E-\mu)$
\\
We remark that $\phi_{\infty}=\zeta_{-}(E-\mu)$, and

\beq
\label{4.13e}
\phi_{\beta}(\widehat{H})=\phi_{\beta}(\widehat{H})\chi(\widehat{H})
\eeq

\noi for some $\chi\in{\cal C}_0^{\infty}(\R)$
because the spectrum
   of $\widehat{H}$ is bounded from below.
\\
But $E\longmapsto\phi_\beta(E)\chi(E)$ is a bounded family of functions in
   ${\cal C}_0^{\infty}({I\hskip-1truemm R})$ for $\beta$ in $]0,+\infty]$.
   Therefore  the $\hbar$ -semiclassical functional calculus can be
applied, yielding an
   asymptotic expansion of the following form~:

\beq
\label{4.14e}
I_{\eta}^-(\mu)\sim\hbar^{-n}\sum_{j\geq 0}c_j\hbar^j
\eeq

\noi uniformly in $\sigma\in ]0,+{\infty}]$, where~:

\beq
\label{4.15e}
c_0=(2\pi)^{-n}\int
f\left(\beta(H(q,p)-\mu)\right)\zeta_{-}(H(q,p)-\mu)\left\{A,B_{t}
\right\}(q,p)dqdp
\eeq

\noi and analogous formulae for $j\geq1$, where we have used the
known result that
the principal symbol of:

\beq
\label{4.16e}
   \widehat{C_t }:= {i\over {\hbar}} [\widehat{A}, \widehat{B}_t]
   \eeq

   \noi is $\left\{A,B_t\right\}$, using Egorov's theorem in the form
given in \cite{BR1}
   for $\vert t \vert < {\gamma}_H Log(1/\hbar)$, and semiclassical calculus.

\bigskip
\noindent
\textbf{Estimate of $I_{\eta}^{+}(\mu)$}~:

\beq
\label{4.17e}
\vert I_{\eta}^+(\mu)\vert\leq C\sum_{j\geq1} \vert f(\beta(E_{j}-\mu))
\zeta_{+}(E_{j}-\mu)\vert
\eeq

\noi where C depends on $\widetilde{g}$ and $A$, $B$, and where
$\left\{E_j\right\}_{j\geq1}$
is the increasing sequence of the eigenvalues of $\widehat{H}$

\beq
\label{4.18e}
\vert I_{\eta}^+(\mu)\vert\leq C\sum_{E_{j}\geq\delta/2}\vert
f(\beta(E_{j}-\mu))\vert
\eeq

\noi We introduce the counting function~:
\\
\\
$N(E):=\sharp\left\{j: E_{j}\leq E\right\}$
\\
\noi Using a ``Lieb-Thirring-like'' estimate, we get~:
\\
\\
$N(E)\leq \gamma {\hbar}^{-n}(1+E)^m$
\\
   We therefore deduce the existence of a positive constant $c$
(depending on $\delta$) such that~:

\beq
\label{4.19e}
\vert I_{\sigma,\infty}^+\vert \leq C {\hbar}^{-n}e^{-c\beta}
\eeq

\noi for any $\hbar\in]0,1]$ and any positive $\sigma$.

\bigskip

\noi \textbf{Estimate of $I_{\eta, \tau_0}^0(\mu)$}\\
\medskip
\noi Recall that~:

   \beq
\label{4.20e}
I_{\eta, \tau_0}^0(\mu)={1/i\hbar}\int_{-\infty}^{+\infty} {\rm tr }
\left(f_{\sigma,\eta,{\tau}_0}
\left(\frac{\widehat{H}-\mu}{\hbar}\right)\zeta_0(\widehat{H}-\mu)
\left[\widehat{A}, \widehat{B}_t\right]\right)\widetilde{g}(t)dt
\eeq

\noi where we have defined:

\beq
\label{4.21e}
f_{\sigma,\eta,\tau_0} = f_{\sigma}*{\eta}*{\theta}_{\tau_0}
\eeq

We want to estimate

\beq
\label{4.22e}
L_{\mu, t}:={1\over{i\hbar}}{\rm tr }\left(f_{\sigma,\eta
,\tau_0}\left(\frac{\widehat{H}-\mu}
{\hbar}\right)
\zeta_0(\widehat{H}-\mu)\left[\widehat{A}, \widehat{B}_t\right]\right)
\eeq

\noi Since~:
\medskip

$f_{\sigma, \eta,\tau_0}(\nu)\rightarrow0$ as $\nu\rightarrow+\infty$
\medskip

\noi we have~:

\beq
\label{4.23e}
f_{\sigma,\eta, \tau_0}\left(\frac{\widehat{H}-\mu}{\hbar}Ì\right)={-1/\hbar}
\int_{-\infty}^{\mu}\left(f'_{\sigma}*\eta*{\theta}_{\tau_0}\right)\left(
\frac{\widehat{H}-\lambda}{\hbar}\right)d\lambda
\eeq

\noi using equ. (\ref{4.12e}) we can write~:

\begin{eqnarray}
\label{4.24e}
L_{\mu,t} = {-1\over{\hbar}}{\rm tr}
\int_{\mu-2\delta}^{\mu}\left(f'_{\sigma}*\eta*{\theta}_{\tau_0}
\right)\left(\frac{\widehat{H}-\lambda}{\hbar}\right)\zeta_0(\widehat{H}-\mu)\widehat{C_t} 
d\lambda\\
-{\hbar}^{-1}{\rm tr}
\int_{-\infty}^{\mu-2\delta}(f'_{\sigma}*\eta*{\theta}_{\tau_0})\left(
\frac{\widehat{H}-\lambda}{\hbar}\right)\zeta_0(\widehat{H}-\mu)\widehat{C_t}d\lambda
\nonumber
\end{eqnarray}
\smallskip

\noi We have thus: $L_{\mu, t}=L_{\mu,t}^{1}+L_{\mu,t}^{2}$, with

\beq
\label{4.25e}
\vert L_{\mu,t}^{2}\vert\leq
C{\hbar}^{-1}\int_{-\infty}^{\mu-2\delta}\sum_{j\geq1}\left\vert
(f'_{\sigma}*\eta*{\theta}_{\tau_0})(\frac{E_{j}-\lambda}{\hbar})\zeta_{0}(E_{j}
-\mu)\right\vert
d\lambda
\eeq

\noi  where $C$ is uniform with respect to $t\in$
Supp${\widetilde{g}}$ and to $\hbar$.
\\
By playing with localization and decay properties, one easily obtains that
   $L_{\mu,t}^2=O({\hbar}^{\infty})$ uniformly with respect to $t\in$
Supp${\widetilde{g}}$,
    and with respect to $\sigma\in]0,{+\infty}[$.
\\
The term $L_{\mu,t}^1$ can be dealt with as in \cite{CR}, using the
fact that for
   $\delta$ small enough, $\lambda$ is non critical for $\widehat{H}$ for every
   $\lambda\in [\mu-2\delta, \mu]$.
Thus $L_{\mu,t}^1$ can be rewritten as:

\beq
\label{4.26e}
L_{\mu,t}^1=-\int_{\mu-2\delta}^{\mu}I(\lambda)d\lambda
\eeq

\noi where~:

\beq
\label{4.27e}
I(\lambda)={i\hbar}^{-1}\int_{-\infty}^{+\infty}ds\frac{s\pi/\sigma}{\sinh{s\pi/
\sigma}}
\widetilde{\theta}(s/ \tau_{0}) \widetilde{\eta}(s)
{\rm tr}\left\{e^{-is(\widehat{H}-\lambda)/\hbar}
\zeta_{0}(\widehat{H}-\mu)\widehat{C_t}\right\}
\eeq

   Then using a coherent states decomposition of the trace as in
\cite{CRR}, we see, using the
   support property of ${\widetilde{\theta}}_{\tau_0}$ that the
dominant contribution in the
   stationary phase theorem comes from $s=0$.
Thus, provided that $2{\tau}_0$ is smaller than the smallest period
of closed orbits on
$\Sigma_{\mu}$, equ. (\ref{4.27e}) provides an asymptotic expansion in $\hbar$,
of the form:

\beq
\label{4.28e}
I(\lambda)={\hbar}^{-n}(C_0(\lambda)+\hbar C_1(\lambda)+....) \qquad
{\rm mod~} O(h^{\infty})
\eeq

\noi which is uniform in $\sigma\in]0, +{\infty}[$, and which can be
further integrated with
respect to $\lambda$ on the interval $[\mu, \mu+2\delta]$, yielding the result.
\bigskip

We shall now give the explicit form of the dominant $O({\hbar}^{-n})$
contribution to
   $I_{\eta}(\mu)$ (equ. (\ref{4.7e})) which comes from the sum of the
contributions of
   $I_{\eta}^{-}(\mu)$ and $I_{\eta, \tau_0}^0(\mu)$; we obtain:

\beq
\label{4.29e}
  I_{\eta}(\mu) =  h^{-n}\widetilde{\eta}(0)\int \widetilde{g}(t)
  \int_{[H\leq{\mu}]} \left\{A, B_t\right\}(q,p)dp dq dt + O(h^{1 - n})
\eeq

We have introduced the correlation in time of $A$ and $B$ on the
energy surface ${\Sigma}_{\mu} =
[H(q,p)=\mu]$ (see (\ref{4.2e}))

\beq
\label{4.30e}
C_{A, B, \mu}(t):=
\int_{[H=\mu]}A B_t\frac{d{\sigma}_{\mu}}{\vert\nabla H \vert}
\eeq

\noi Let $\varphi$ be a $ {\cal C}^{\infty} $ function with compact
support contained in $] -\infty, \mu+\delta]$. We have~:

\beq
\label{4.31e}
\int \{ A,B_t\}\varphi(H(q,p))dpdq=\int\{A\varphi(H),B_t\}dqdp-\int
A\{\varphi(H), B_t\}dqdp
\eeq

\noi where the integration is over the full phase space  ${\R}^{2n}$.
$B\varphi(H)$ being a ${\cal C}_{0}^{\infty}(\R^{2n})$
function of $(q,p)$,
we get by integration by part that~:

$$\int\{A\varphi(H), B_t\}dqdp=0$$

\noi Moreover

$$
\int \{A\varphi(H),B_t\}dqdp=\int\{H, B_t\}\varphi'(H)A dqdp=-{d\over{dt}}
\int A B_{t}\varphi'(H)dqdp
$$

\noi We now let $\varphi$ tend to $\1_{]-{\infty},\mu]}$, and find~:

$$\int_{[H\leq\mu]}\{A, B_t\} dqdp={d\over{dt}}\left(\int_{[H=\mu]}
A.B_t \frac{d \sigma_{\mu}}
{\vert\nabla H\vert}\right)
$$

\noi  Therefore the dominant term of $I_{\eta}(\mu)$ is given by~:

$$
  I_{\eta}(\mu) =  h^{-n}\widetilde{\eta}(0)\int C_{A, B,
  \mu}(t)\widetilde{g}~'(t) dt + O(h^{1 - n})
$$

This completes the proof for the first term of the asymptotic
expansion in Theorem (4.1).
\bigskip

\noi \textbf{Estimate of $I_{\sigma, \tau,\tau_{0}}^{osc}(\mu)$}
\\
We have~:

\beq
\label{4.32e}
   I_{\eta,\tau_{0}}^{osc}(\mu)=h^{-1}\int dt \widetilde{g}(t)\int \frac{1}{s}ds
   \frac{\pi s/\sigma}{\sinh{\pi s/\sigma}}\eta_{\theta, \tau_0}(s)
{\rm  tr}\left\{\zeta_0(\widehat{H}-\mu)e^{-is(\widehat{H}-\mu)/\hbar}\widehat{C_t}\right\}
\eeq

\noi where we have used the following notation~:

$$
\eta_{\theta, \tau_0}(s):=
\widetilde{\eta}(s)\left(1-\widetilde{\theta}_{\tau_0}(s)\right)
$$

\noi Again we proceed as in \cite{CRR} by a ``Gutzwiller type''
estimate for the integral over $s$
since the support of $\eta_{\theta, \tau_0}(s)$ doesn't contain
$s=0$, but will only contribute
by a finite number of closed classical orbits which we denote by
$\gamma$. Furthermore due to
   the support properties of $\widetilde{\theta}$, it is clear that
    $\eta_{\theta, \tau_0}(T_{\gamma})  = \eta(T_{\gamma})$

\medskip

Using the \textbf{Gutzwiller assumption}, and the compact integration
support in variable $t$,
   we obtain the following asymptotic expansion of
$I_{\eta,\tau_{0}}^{osc}(\mu)$, which
   is uniform in the parameter $\sigma\in ]0, +\infty[$:

\bea
\label{4.33e}
I_{\eta,\tau_{0}}^{osc}(\mu) = h^{-1}\int dt
\widetilde{g}(t)\sum_{\gamma\in\Sigma_{\mu}}
\frac{\pi}{\sigma\sinh{\pi T_{\gamma}/\sigma}}
\frac{e^{iS_{\gamma}/\hbar +i\nu_{\gamma}\pi/2}}
{\vert \det(1-P_{\gamma})\vert^{1/2}}\times\\
\times\left(\sum_{k}\eta(T_{\gamma})c_{\gamma^*,k}e^{2i\pi kt/T_{\gamma}^*}+
\sum_{j\geq1}h^{j}\nu_{j, \gamma}(t)\right)
\nn
\eea

\noi This yields the following contribution of oscillating terms  to
$\chi_{A, B}(s,
\omega)$, in the distribution sense,
    uniformly in $\sigma\in ]0,+{\infty}[$ ~:

\bea
\label{4.34e}
  \chi_{A, B, osc} =   \sum_{\gamma :T_{\gamma}\not=0}\frac{\pi
e^{i(S_{\gamma}/\hbar+\nu_{\gamma}\pi/2)}}
{\hbar\sigma \sinh {(\pi T_{\gamma}/{\sigma})}\vert
det(1-P_{\gamma})\vert ^{1/2}}\times \nonumber \\
\times\left(\delta_{T_{\gamma}}(s)
\otimes\sum_{k}c_{\gamma*,k}\delta(\omega-\frac{2k\pi}{T_{\gamma*}}
) +\sum_{j\geq1} {\hbar}^j\nu_{j,\gamma}(s,\omega)\right)
\eea

\noi where $\nu_{j,\gamma}(s,\omega)$ are distributions supported
in $\{T_{\gamma}\}\times\R$
\QED

\mysection{Concluding remarks}

  Theorem 4.1 is an extension of the well known Gutzwiller trace
  formulae  for the spectral density of energy levels. The main
  difference is that  here there are  two real variables instead of one  because
in the ``dynamical susceptibility''  time and energy variables are mixed
up in an intricated way. So  we can put in a mathematical rigorous
shape the main result of the  paper \cite{GJ}. \\ \noi
	As in the Gutzwiller trace formulae  our formulae in Theorem 
4.1 gives a semiclassical expansion
  with three different terms: the first line gives a regular expansion
in $h$,  which is the contribution of the period 0 of the classical flow;
the second line is an oscillating part coming from the
  contributions of the non zero periods of the classical flow;
the third line is the error term  depending on the test functions considered.

So far we have shown that a semiclassical expansion, in the linear response
theory, can be obtained for
   a regularized version of the ``dynamical susceptibility'', i.e in a
suitable distributional sense.
    The same is obviously true for the linear response function
$J_L(t)$ defined by (\ref {3.11e}),
    as we shall establish now. \\

    Formally, if $\varphi$ is a ${\mathcal C}_0^\infty (\R)$ function, we
have, in distributional sense:
$$<J_L,\varphi> =   \int \widetilde{k_1}( \omega) \chi_{A,B}( \omega) d
\omega$$

\noi where $k_1(u):= \Theta(u) k(u)$, \\
\noi ($\Theta$ being the Heavyside function)\\
\noi and $k(u):= \int \varphi(s+u) F(s) ds$ \\
\noi However $\chi_{A,B}$ is only well defined mathematically as a
semiclassical expansion in a
``regularized'' form:

$$\chi_{A,B,\eta}:= \int \widetilde{\eta}(s) \chi_{A,B} (s, \omega) ds$$

\noi where $\widetilde{\eta}$ is in ${\mathcal C}_0^\infty (\R)$.\\

\noi Similarly, a ``regularized form'' of $J_L$ can be defined as:

$$J_{L,\eta}(t):= \int \widetilde{\eta} (s) J_L(t,s) ds$$

\noi in the following sense:

$$<J_{L,\eta}, \varphi> =   \int \widetilde {k_1} (\omega) \chi_{A,B,
\eta} (s, \omega) d \omega$$
So we have
$$ <J_{L,\eta}, \varphi> = \int \widetilde {\eta} (s) \widetilde {k_1} (\omega)
\chi_{A,B} (s, \omega)ds d \omega $$

\noi It is not hard to see, using the definition of $k_1$ that
$\widetilde{\eta}(s)
\widetilde {k_1} (\omega) \in {\mathcal K}_{\alpha- \varepsilon}$ for
any $\varepsilon >0$,
   so that our theorem applies. For example  we can compute the leading
   term~:
   $$
   <J_{L,\eta},\varphi> =
    -2\pi h^{-n}\tilde\eta(0)\int_0^{+\infty}duC^\prime_{A, B}(u)\int_{\R}ds
\varphi(s +
   u)F(s) + O(h^{1-n})
   $$

\bibliographystyle{amsalpha}

\end{document}